# Thermally Assisted Current-Driven Bistable Precessional Regimes in Asymmetric Spin Valves

M. Gmitra[1,2] and J. Barnaś[2]

[1]*Institute of Physics, P. J. Šafárik University, Park Angelinum 9, 040 01 Košice, Slovak Republic*
[2]*Department of Physics, Adam Mickiewicz University, Umultowska 85, 61-614 Poznań, Poland*

Spin-transfer torque in asymmetric spin valves can destabilize both parallel and antiparallel configurations and can lead to precessional modes also in the absence of an external magnetic field. We find a bistable precessional regime in such systems and show that thermal fluctuations can excite transitions (telegraph noise) between the corresponding oscillatory regimes that are well separated by irreversible paths at low temperatures. Because of the thermally induced transitions, the frequency of the resulting current-driven oscillations is different from that obtained at very low temperatures. We also show that the power spectrum in the bistable region is dominated by the out-of-plane oscillatory mode.

*Introduction.*—The phenomenon of current-induced magnetic switching of thin ferromagnetic films due to spin transfer from conduction electrons to localized moments has been extensively studied in recent literature both theoretically [1] and experimentally [2]. In some conditions the spin current can cause a transition to steady precessional modes, where energy is pumped from conduction electrons to localized magnetic moments [3–6]. In Co/Cu/Co nanopillars, the steady precessions have been observed for both an external magnetic field and current density exceeding relevant critical values [4,5].

In our earlier papers [7–9] we have predicted that current-induced microwave oscillations in certain asymmetric spin valves can exist without external magnetic field due to a nonstandard angular dependence of the spin-transfer torque in the diffusive transport regime. This prediction has been recently confirmed experimentally on Co/Cu/Py nanopillars [10]. In this Letter we consider temperature effects on spin switching and spin dynamics and report new features. First of all, we found a bistable zero-field precessional regime, in which the thermal energy can induce telegraph noise due to jumps between two oscillatory regimes. Second, the calculated power spectra in the bistable regime clearly show a blueshift of microwave oscillations, in agreement with experiment [10].

*Model and description.*—We consider the spin valve IrMn/Co(6)/Ru(2)/Co(4)/Cu(8)/Py(4)/Cu [see Fig. 1(a)] in the diffusive transport regime (the numbers in parentheses are the layer thicknesses in nm) [8]. The system consists of pinned and reference cobalt layers [Co(6) and Co(4), respectively] separated by a ruthenium layer, copper spacer layer, and sensing permalloy layer. The P and AP configurations refer to the relative alignment of magnetic moments of the reference and sensing layers. Strong antiferromagnetic coupling is assumed between the Co layers across the Ru layer [11]. The Co layer pinned *via* exchange biasing to IrMn as well as the reference Co layer do not undergo dynamics in current densities of interest.

Magnetic dynamics of the sensing layer is described by the Landau-Lifshitz-Gilbert equation,

$$\frac{d\hat{s}}{dt} = -|\gamma_g|\mu_0 \hat{s} \times \boldsymbol{H}_{\text{eff}} - \alpha \hat{s} \times \frac{d\hat{s}}{dt} + \frac{|\gamma_g|}{M_s d}\boldsymbol{\tau}, \quad (1)$$

where $\hat{s}$ is the unit vector along the spin moment of the sensing layer, $\gamma_g$ the gyromagnetic ratio, $\mu_0$ the magnetic vacuum permeability, $d$ thickness of the sensing layer, $\boldsymbol{H}_{\text{eff}}$ an effective magnetic field for the sensing layer, $\alpha$ the damping parameter, and $M_s$ the saturation magnetization. The effective field $\boldsymbol{H}_{\text{eff}}$ includes the uniaxial magnetic anisotropy field $H_a$ and the demagnetization field $\boldsymbol{D}$; $\boldsymbol{H}_{\text{eff}} = -H_a(\hat{s}\cdot\hat{e}_z)\hat{e}_z + \boldsymbol{D}$, where $\hat{e}_z$ is the unit vector along the $z$ axis (in-plane), and $\boldsymbol{D} = M_s N\hat{s}/2$, where $N$ is the diagonal matrix with $\text{Tr}\{N\} = 1$. We assume the demagnetization factors for a flat ellipsoid [12]. The last term in Eq. (1) stands for the torque due to spin transfer, $\boldsymbol{\tau} = \boldsymbol{\tau}_\theta + \boldsymbol{\tau}_\varphi$, with $\boldsymbol{\tau}_\theta = aI\hat{s}\times(\hat{s}\times\hat{S}) = \tau_\theta\hat{e}_\theta$, and $\boldsymbol{\tau}_\varphi = bI\hat{s}\times\hat{S} = \tau_\varphi\hat{e}_\varphi$. Here, $\hat{S}$ is the unit vector along the spin moment of the reference magnetic layer ($\hat{S} = \hat{e}_z$), while $\hat{e}_\theta$ and $\hat{e}_\varphi$ are the unit vectors of a coordinate system associated with the polar $\theta$ and azimuthal $\varphi$ angles which describe orientation of the vector $\hat{s}$. The current $I$ is positive when it flows from the sensing layer towards the reference one. The parameters $a$ and $b$ depend on the angle between spin moments of the sensing and reference layers and have been calculated in the diffusive transport regime [9].

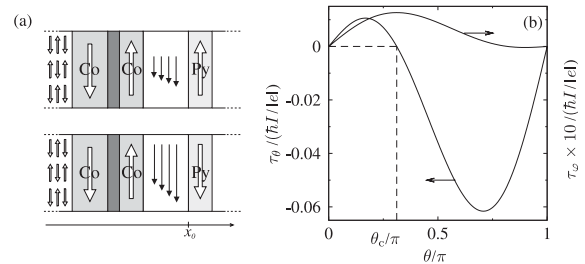

FIG. 1. (a) Schematic structure of the nanopillar and spin accumulation in thin spacer layer for both collinear configurations, and for electrons flowing along the axis $x$ ($I > 0$). (b) Angular dependence of the torques $\tau_\varphi$ and $\tau_\theta$ acting on the Py layer in the IrMn/Co(6)/Ru(2)/Co(4)/Cu(8)/Py(4)/Cu spin valve. For the Py layer we assumed: $M_s = 10.053$ kOe, $H_a = 2.51$ Oe, $\alpha = 0.01$, and the demagnetization matrix $N$ is taken to model ellipsoid of the thickness $d = 4$ nm and radii 140 nm and 70 nm. The other parameters are (Co) bulk resistivity $\rho^* = 5.1$ $\mu\Omega$ cm, spin asymmetry factor $\beta = 0.51$, and spin-flip length $l_{\text{sf}} = 60$ nm; (Py) $\rho^* = 16$ $\mu\Omega$ cm, $\beta = 0.77$, $l_{\text{sf}} = 5.5$ nm; (Cu) $\rho^* = 0.5$ $\mu\Omega$ cm, $l_{\text{sf}} = 1000$ nm; (Ru) $\rho^* = 9.5$ $\mu\Omega$ cm, $l_{\text{sf}} = 14$ nm; (IrMn) $\rho^* = 150$ $\mu\Omega$ cm, $l_{\text{sf}} = 1$ nm. In turn, for the Co/Cu interface we assume the interfacial resistance per unit square $R^* = 0.52 \times 10^{-15}$ $\Omega$ m$^2$, interface spin asymmetry factor $\gamma = 0.76$, and the mixing conductances Re$G_{\uparrow\downarrow} = 0.542 \times 10^{15}$ $\Omega^{-1}$ m$^{-2}$ and Im$G_{\uparrow\downarrow} = 0.016 \times 10^{15}$ $\Omega^{-1}$ m$^{-2}$; for the Py/Cu interfaces we assume $R^* = 0.5 \times 10^{-15}$ $\Omega$ m$^2$, $\gamma = 0.7$, Re$G_{\uparrow\downarrow} = 0.39 \times 10^{15}$ $\Omega^{-1}$ m$^{-2}$, and Im$G_{\uparrow\downarrow} = 0.012 \times 10^{15}$ $\Omega^{-1}$ m$^{-2}$; for IrMn/Co we take $R^* = 0.5 \times 10^{-15}$ $\Omega$ m$^2$, $\gamma = 0.1$; and for Co/Ru we assume $R^* = 0.5 \times 10^{-15}$ $\Omega$ m$^2$, $\gamma = -0.2$.

In Fig. 1(b) we show the angular dependence of the torques $\tau_\varphi$ and $\tau_\theta$ acting on Py layer in the studied valve. The plot shows that $\tau_\theta$ vanishes at $\theta = \theta_c \neq 0, \pi$, which indicates that both P and AP configurations can be unstable for a positive current exceeding a certain critical value. To observe such nonstandard behavior of $\tau_\theta$ one needs: (i) the spin-flip length in the sensing layer significantly shorter and/or (ii) the asymmetry in bulk resistivities of the sensing layer significantly larger than those in the reference layer. This leads to an inverse spin accumulation in the spacer layer, see Fig. 1(a), due to significant backscattering in the spin-down channel at the Cu/Py interface for both collinear configurations.

*Zero-temperature limit.*—Figure 2(a) shows the reduced magnetoresistance [13], $r = [1 - \cos^2(\theta/2)]/[1 + \cos^2(\theta/2)]$, as a function of the current density. In asymmetric valves the angular magnetoresistance can not be a monotonous function of the angle [14]. The arrows indicate direction of the current change. The initial AP configuration is stable for $I < 0.1I_0$, where $I_0 = 10^8$ A/cm$^2$. Further increase in current drives the system *via* the in-plane oscillatory modes near the $-z$ axis (IP$_-$ modes) to the P state. The arc length of the steady IP$_-$ orbits increases with increasing $I$ [Fig. 2(e)]. The angular speed changes then very little, so the corresponding oscillation frequency $\omega_0$ decreases (redshift), as shown in Fig. 2(b). The P state is stable up to $I = 0.36I_0$, and then the system is driven to the in-plane oscillations near the $+z$ axis (IP$_+$ modes). From the ratio of the critical threshold currents [8] one can conclude on the parameter $a$, $-a(\theta = \pi)/a(\theta = 0) \simeq 3.66$. Because of symmetry of the considered $\boldsymbol{H}_{\text{eff}}$, the steady IP$_+$ oscillations show redshift similarly as the IP$_-$ modes. The IP$_+$ regime is stable up to $I = 1.95I_0$, where the system is switched to high-resistance static state (HSS). The nonzero $\tau_\varphi$ at $\theta \to \theta_c$,

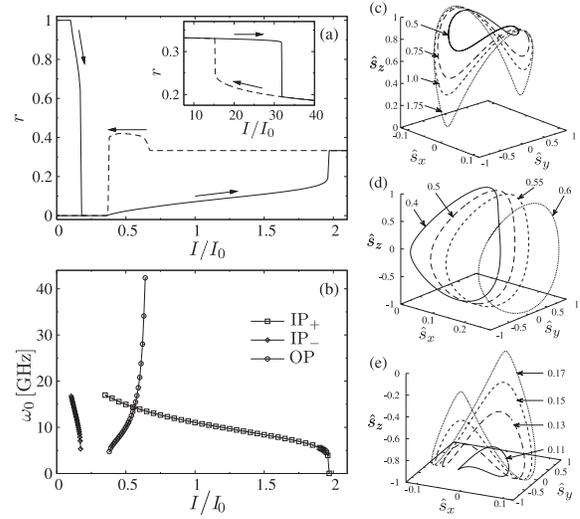

FIG. 2. (a) Current-driven hysteretic behavior of the magnetoresistance in zero field, scanned with increasing (solid line) and decreasing (dashed line) current with the sweeping rate $1.2\times 10^4$ A/cm$^2$ s. (b) Fundamental frequency $\omega_0$ of the magnetoresistance oscillations calculated as a function of the current. The in-plane steady state orbits (IP$_+$) near the $+z$ axis (c), the out-of-plane (OP) orbits (d), and the in-plane orbits (IP$_-$) near the $-z$ axis (e) are presented for indicated values of the current. The other parameters as in Fig. 1

$$\tau_\varphi = -\frac{\hbar}{e^2}\left[\text{Re}\,G_{\uparrow\downarrow} + \frac{(\text{Im}\,G_{\uparrow\downarrow})^2}{\text{Re}\,G_{\uparrow\downarrow}}\right] \\ \times (g_x\sin\varphi - g_y\cos\varphi)\Big|_{\substack{\theta\to\theta_c \\ x\to x_0}} \qquad (2)$$

assists in the transition from IP$_+$ to HSS state and thus reduces the critical current for switching. Here, the $G_{\uparrow\downarrow}$ is the mixing conductance for the active Cu/Py interface, while $g_x$ and $g_y$ are components of the corresponding spin accumulation. When Im $G_{\uparrow\downarrow}$ vanishes for the Co/Cu interface, one finds $\tau_\varphi = \tau_\theta = 0$ at $\theta = \theta_c$, and the region of IP$_+$ modes extends to high current densities.

The HSS states are stable up to a large current density and give rise to the current-driven hysteresis between HSS states and static states (SS) of lower resistance, see the inset to Fig. 2(a). Switching between the steady states is driven *via* transient regimes, where spin transfer pumps energy to the system ($T_+$) or dissipates energy ($T_-$), respectively. The hysteresis can be formally written as a sequence of transitions AP $\to$ IP$_-$ $\to$ P $\to$ IP$_+$ $\to$ $T_+$ $\to$ HSS $\to$ $T_-$ $\to$ SS for increasing current and SS $\to$ $T_+$ $\to$ HSS $\to$ $T_-$ $\to$ OP $\to$ P for decreasing current, where OP stands for the steady out-of-plane oscillations, see Fig. 2(d). The OP modes exist roughly in the region $0.4 < I/I_0 < 0.6$. The corresponding vector $\hat{s}$ processes mostly in the out-of-layer plane with the angular velocity roughly proportional to $|\gamma_g|D_x$. With decreasing $I$, balance be-tween the energy pumped *via* spin transfer and sustained Gilbert dissipation leads to reduction of the out-of-plane component of $\hat{s}$, and consequently $\omega_0$ decreases with decreasing current (blueshift), see Fig. 2(b). However, both the IP$_+$ and OP regimes are well separated by irreversible paths, as follows from the presence of hysteresis.

*Thermal activation.*—Finite temperature results in a finite probability for thermally activated switching, which may play a significant role, particularly in nanoscale systems. To study the temperature effects on spin valve behavior, we model the thermal fluctuations by adding a Langevin random field $\mathbf{H}_r$ to the effective field in Eq. (1). The spin-transfer torque comes from conduction electrons whose transport properties are less affected by thermal fluctuations since the Fermi level is usually much higher than the thermal energy [15]. Thus, the leading fluctuation term in the spin-transfer torque is due to its dependence on the thermal magnetization fluctuations.

The field $\mathbf{H}_r$ is a fluctuating random field whose statistical properties are defined as $\langle H_r^i(t) \rangle = 0$ and $\langle H_r^i(t) H_r^j(t') \rangle = 2D\delta_{ij}\delta(t-t')$, where $i$ and $j$ are Cartesian indices, and $\langle \ldots \rangle$ denotes an average over all realizations of the fluctuating field. According to the fluctuation-dissipation relation [16], the parameter $D$ represents the strength of thermal fluctuations, $D = \alpha k_B T / \mu_0^2 |\gamma_g| M_s V$, where $V$ is volume of the sensing layer. However, spin transfer can dissipate or pump energy into the system. To redeem the concept of stationary process, one can introduce an effective temperature [15] or an effective energy barrier [17,18], which depend on spin transfer. The effective temperature should be understood in terms of a stationary solution for the probability density. Furthermore, the critical current rescales by the factor of $1 - (k_B T/E) \ln(\tau/\tau_0)$ [18] due to thermal activation, where $E$ is the actual energy barrier, whereas $\tau$ and $\tau_0$ are the thermal activation lifetime and inverse attempt frequency, respectively. Thus, in the low current regime or in subcritical region ($I < I_{c0}$, where $I_{c0}$ is the zero-temperature threshold current density), finite temperature gives rise to a nonvanishing probability for thermally activated switching.

The reduced magnetoresistance averaged over several hundreds of realizations, shown in Fig. 3(a) as a function of current, reveals vanishing hysteretic behavior. The double peak structure in the power spectra for $0.4 < I/I_0 < 0.55$, see Fig. 3(b), indicates that the vanishing hysteresis is a result of "telegraph" jumps between the IP$_+$ and OP regimes, see Fig. 3(c). Both IP$_+$ and OP oscillatory regimes are well resolved for $I = 0.4 I_0$ due to well-separated frequencies of both modes, as follows from Fig. 2(b).

One should note that the peaks in power spectra are relatively broad and one of them survives even for vanishing current. When the system settles in the P state, the sustained thermal fluctuations and the subsequent relaxation back to the P state induce low frequency noise and a broad peak at 16.4 GHz, with the full linewidth at half maximum (FWHM) of 0.9 GHz. This is due to thermal deflection of the magnetic moment about the equilibrium trajectory that results from competition between the energy dissipation and thermal activation.

Current-driven dynamical modes lead to narrower linewidths. Moreover, the spin transfer in real systems can generate linewidths even narrower than those expected in the macrospin model [19]. Apart from the thermal deflection and thermal fluctuations that displace the moment both along and transverse to the equilibrium trajectory, the thermally activated transitions between different dynamical modes lead to linewidths growing with increasing current. Figure 3(d) shows the temperature dependence of the FWHM of the IP$_+$ modes ($T < 160$ K) for $I = 0.5 I_0$. The FWHM increases as $\propto \sqrt{T}$ up to $T = 160$ K, see the solid line which is a fit to numerical data. The discontinuous jump at $T = 160$ K reveals the bistability and onset of the transitions from IP$_+$ to OP regimes.

The frequencies of the IP$_+$ and OP magnetoresistance oscillations, obtained from the power spectra at room temperature, are shown in Fig. 3(e). These frequencies for $I = 0.5I_0$ become closer to each other when the temperature increases, see Fig. 2(b) and 3(e). The presented spectra reveal increasing power of the OP regime, which dominates the (decreasing) power of IP$_+$ regime for $I > 0.44I_0$. The broad peak due to the OP modes screens then the narrower peak resulting from the IP$_+$ regime. Thus, by analyzing power spectra of the magnetoresistance one can conclude about the OP regime only [10].

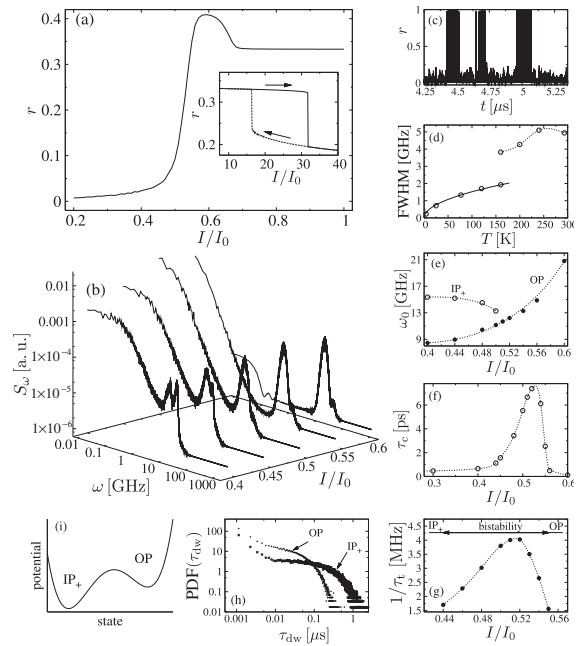

FIG. 3. (a) Room temperature behavior of the current-driven magnetoresistance, scanned with the sweeping current rate $1.2 \times 10^4$ A/cm$^2$ s. (b) Power spectra of the magnetoresistance for current densities as indicated. (c) Telegraph noise of magnetoresistance for $I = 0.5 I_0$. (d) FWHM as a function of temperature for $I = 0.5 I_0$. The solid line is the fit to the data. (e) Frequency $\omega_0$ of the magnetoresistance oscillations obtained from the power spectra. (f) Current-driven autocorrelation time of magnetoresistance and (g) transition rate between the IP$_+$ and OP regimes. (h) Probability distribution function of dwell times for $I = 0.5 I_0$. (i) Schematic plot of double-well potential. The other parameters as in Fig. 1.

The relevant quantity of the analysis is the magnetoresistance autocorrelation function, $C(t) = \langle r(0)r(t)\rangle$, and the autocorrelation time $\tau_c = \int_0^\infty C(t)dt$. Variation of $\tau_c$ with the current density, shown in Fig. 3(f), reveals a peak in the region where the IP$_+$ and OP regimes are strongly correlated due to thermal excitations. The current dependence of the transition rates between the two regimes, $1/\tau_t$, is also peaked in this region, see Fig. 3(g).

We have analyzed the bistability by accumulating the probability distribution function (PDF) of the dwell times. The dwell time $\tau_{\rm dw}^{\rm IP_+}$ ($\tau_{\rm dw}^{\rm OP}$) is the time the system is in the IP$_+$ (OP) regime. Figure 3(h) shows the PDF of the dwell times for $I = 0.5I_0$. The exponential tails are related to the Poisson process, where thermally induced hopping *via* the potential barrier and decay into the steady oscillatory regime take place at a rate described by the Arrhenius law, $\propto \exp(-E/k_B T)$, where $E$ is the energy barrier modified by the spin transfer. The $\tau_{\rm dw}^{\rm IP_+}$ for $I = 0.5I_0$ decays exponentially 6 times slower than $\tau_{\rm dw}^{\rm OP}$. Thus, the system settles mostly in the IP$_+$ state and rarely in the OP regime. With increasing current, the situation becomes reversed. On the other hand, at short times the PDF is proportional to $\tau_{\rm dw}^{-3/2}$, which is related to over-barrier processes. The power-law dependence can be identified as PDF of the first-passage time of a Brownian particle attempting the barrier in double-well potential. A simple model of Brownian particle in a double-well potential [see Fig. 3(i)] provides physical insight into the mechanism underlying the bistability. A nonzero temperature decreases the effective barrier and connects both the wells (steady states). The current, in turn, modifies asymmetry of the potential. Periodical change in current around the value where the bistability occurs (with the frequency $\omega \ll \omega_0$) can induce controlled jumps between both oscillatory regimes, and thus increases signal to noise ratio known in bistable systems as stochastic resonance [20].

In conclusion, we have found a current-induced bistable precessional region in asymmetric spin valves with antiferromagnetically coupled pinned and reference layers. The bistability also has been shown to have a significant influence on transport characteristics at room temperature. The main advantage of systems with coupled pinned and reference layers over the conventional nanopillars consists in a strong suppression of the dipolar field acting on the sensing layer. The presence of sufficiently large dipolar field could lead to current-driven transition to OP regime and suppression of the bistability, so in the absence of external magnetic field one could identify only the OP regime [10]. We believe that adopting the ideas of stochastic resonance will help to identify experimentally the precessional bistability.

We thank D. Horváth for useful discussions. This work is partly supported by Slovak Ministry of Education as a research project MVTS No. POL/SR/UPJS07 and Slovak Grant Agency VEGA No. 1/2009/05. As part of the European Science Foundation EUROCORES Programme SPINTRA, it was also supported by funds from the Polish Ministry of Science and Higher Education as a research project in years 2006–2009 and the EC Sixth Framework Programme, under Contract No. ERAS-CT-2003-980409.